\documentclass[conference]{IEEEtran}

\usepackage[utf8]{inputenc}
\usepackage[pdftex]{color,graphicx}
\usepackage[caption=false]{subfig}
\usepackage[noadjust,nocompress]{cite}
\usepackage{amsfonts}
\usepackage{tikz}
\usepackage{xspace}

\usepackage{pdfpages}
\usepackage{enumitem}

\usepackage[pdftex]{hyperref}
\hypersetup{
	pdfsubject = {Stream-based State-Machine Replication},
	pdftitle = {Stream-based State-Machine Replication},
	pdfauthor = {Laura Lawniczak and Tobias Distler},
	pdfkeywords = {State-machine replication, fault tolerance, consensus, scalability, data-stream processing},
	pdfborder = {0 0 0}
}

\newcommand{\system}[0]{\textsc{Tara}\xspace}

\newcommand{\headline}[1]{\vspace{1mm}\noindent\textbf{\textit{#1.}}~}

\newcommand{\numlabel}[1]{\begin{minipage}{2.8mm}\vspace{.2mm}\begin{tikzpicture}\node[draw, circle, minimum width=8, font=\scriptsize\sffamily, inner sep=0] {#1};\end{tikzpicture}\end{minipage}}

\newcommand{\msg}[2]{$\langle \textsc{#1}, #2 \rangle$\xspace}

\title{\vspace{-.5em}Stream-based State-Machine Replication\\\vspace{-.5em}{\normalsize\textbf{(Extended Version)}}}

\author{
	\IEEEauthorblockN{Laura Lawniczak and Tobias Distler}
	\IEEEauthorblockA{
		\textit{Friedrich-Alexander University Erlangen-N\"urnberg (FAU)}\\
		Email: \{lawniczak,distler\}@cs.fau.de
	}
}

\begin{document}

\maketitle

\thispagestyle{plain}
\pagestyle{plain}


\begin{abstract}
Developing state-machine replication protocols for practical use is a complex and labor-intensive process because of the myriad of essential tasks~(e.g.,~deployment, communication, recovery) that need to be taken into account in an implementation. In this paper, we show how this problem can be addressed with stream-based replication, a novel approach that implements a replication protocol as application on top of a data-stream processing framework. With such framework already handling most essential tasks and furthermore providing means for debugging and monitoring, this technique has the key benefit of significantly minimizing overhead for both programmers as well as system operators. Our first stream-based protocol \system tolerates crashes and comprises full-fledged mechanisms for request \linebreak handling, checkpointing, and view changes. Still, \system's prototype implementation, which is based on Twitter's Heron framework, consists of fewer than 1,500~lines of application-level code.%
\end{abstract}




\section{Introduction}

\vspace{-.05mm}

State-machine replication protocols such as Paxos~\cite{lamport98part} or Raft~\cite{ongaro14search} represent corner stones of many dependable services in production by enabling a system to tolerate crashes of participating processes. Unfortunately, implementing and operating these protocols usually is a difficult and time-con\-su\-ming undertaking; not only due to the protocols' inherent complexity~\cite{chandra07paxos,kirsch08paxos}, but especially because there is a multitude of tasks that need to be taken care of in practice. Among other things, replicas for example must be installed and started on different servers, network connections have to be set up and maintained between nodes, exchanged messages need to be serialized and delivered to their intended receivers, and failed processes should be detected and recovered. All previously mentioned tasks have in common that they typically are not part of the replication-protocol logic and therefore further add to the complexity of the overall implementation. Nevertheless, since the tasks are essential for the execution of a replication protocol, they must be handled in some form or the other, for example by integrating external libraries (if possible) or by implementing them manually. Apart from complicating protocol development in general, this particularly makes it difficult to quickly create prototypes for testing new ideas.

In this paper, we present an approach that addresses these problems by implementing a state-machine replication protocol as application on top of a stream processing framework (e.g.,~Heron~\cite{kulkarni15twitter}, Storm~\cite{toshniwal14storm}, or Flink~\cite{carbone15apache}). Our choice of stream processing frameworks as underlying platform is motivated by several reasons: (1)~With stream processing applications being widely used to analyze data, the frameworks are often already available and operational in many data centers. (2)~Taking care of tasks such as distributed deployment, communication, or the automated recovery of crashed application nodes, the frameworks provide many features that for replication protocols so far had to be specifically integrated. (3)~Stream processing frameworks typically offer built-in support for scalability, which with our approach can be leveraged to improve the performance of replication protocols. (4)~The frameworks are usually equipped with a logging infrastructure, means to collect runtime metrics~(e.g.,~throughput), as well as graphical user interfaces, which each greatly facilitate the development and management of replication-protocol implementations.

Stream processing applications are designed as sets of processing nodes through which data tuples flow along the edges of directed acyclic graphs. This data-oriented perspective stands in sharp contrast with the replica-oriented perspective commonly used to specify state-machine replication protocols, where a small number of replicas repeatedly exchange messages in multiple phases~\cite{kirsch08paxos,ongaro14search}. To show how to bridge this gap we present \system, a \emph{stream-based} replication protocol that has been specifically tailored to run in conjunction with stream processing frameworks. In order to be broadly applicable, \system requires no modifications to the underlying platform and makes only weak assumptions about the services a framework provides with regard to deployment and communication. In particular, there is no need for the framework to implement consensus or replication-based fault tolerance at lower layers.

We implemented \system based on Heron, a stream processing framework developed by Twitter for use in production. Thanks to leveraging Heron, \system's code base for request handling, checkpoint-based garbage collection, and view changes is about two thirds smaller than the implementation of the same tasks in the widely used replication library BFT-SMaRt~\cite{bessani14state}.

In summary, this paper makes the following contributions: (1)~It proposes an approach that facilitates the development and operation of replication protocols by implementing them as applications on top of stream processing frameworks. (2)~It presents the design and implementation of \system, the first stream-based replication protocol. (3)~It uses \system as an example to illustrate how stream-based protocols can leverage the underlying framework to achieve parallelism. (4)~It evaluates \system in the context of a coordination-service application.

The remainder is structured as follows:
Section~\ref{sec:background} introduces background on replication protocols and stream processing frameworks. Section~\ref{sec:approach} describes \system with Section~\ref{sec:parallelization} adding a parallelized version and Section~\ref{sec:implementation} offering implementation details. After that, Section~\ref{sec:evaluation} evaluates \system. Finally, Sections~\ref{sec:related} and \ref{sec:conclusion} discuss related work and conclude.


\section{Background and Problem Statement}
\label{sec:background}

This section provides background on replication protocols and stream processing frameworks and discusses the benefits and difficulties of integrating the former into the latter.

\subsection{Replication Protocols}
\label{sec:background-replication}

\begin{figure}[b!]
	\includegraphics{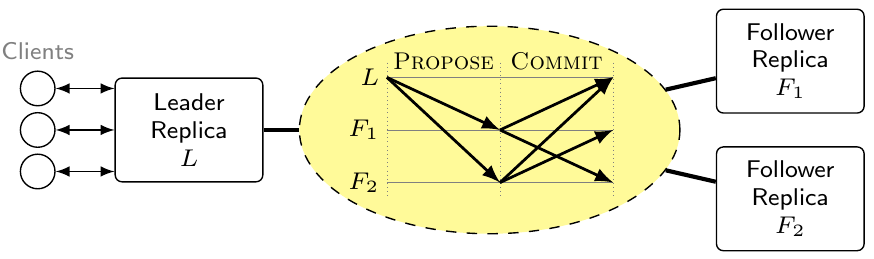}
	\vspace{-6mm}
	\caption{Basic architecture of replicated systems}
	\label{fig:replication}
\end{figure}

State-machine replication protocols~\cite{lamport98part,ongaro14search} tolerate server crashes by modeling a system as a collection of replicas that each maintain an instance of the application state. To keep their state copies consistent, the replicas repeatedly execute a consensus protocol to agree on a common sequence in which to process newly incoming client requests. As illustrated in Figure~\ref{fig:replication}, many protocols for this purpose assign replicas with different roles. A leader replica proposes a specific sequence number for a request, whereas its follower replicas are responsible for committing the sequence-number assignment.

In addition to consensus, replication protocols typically comprise further mechanisms that are essential for the well-functioning of a replicated system. Among other things, this specifically includes sub-protocols for checkpointing and view changes. While periodic checkpoints allow replicas to garbage collect consensus messages, a view-change mechanism enables a system to elect a new leader in case the old one is no longer able to fulfill its duties~(e.g.,~due to having crashed).


\subsection{Stream Processing Frameworks}

\begin{figure}
	\includegraphics{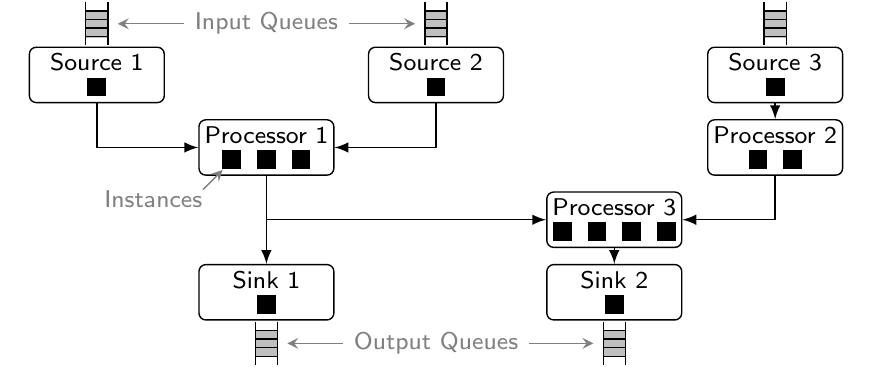}
	\vspace{-6mm}
	\caption{Basic architecture of stream processing applications}
	\label{fig:data-stream-processing}
\end{figure}

Stream processing frameworks such as Heron~\cite{kulkarni15twitter}, Storm~\cite{toshniwal14storm}, or Flink~\cite{carbone15apache} are widely used in production, especially for scenarios in which new data~(e.g., recently posted tweets~\cite{kulkarni15twitter}) needs to be quickly analyzed once it becomes available. As shown in Figure~\ref{fig:data-stream-processing}, stream processing applications are implemented as a set of nodes that are organized in a directed acyclic graph through which information flows from one or more source nodes to one or more sink nodes. Each node in the graph (if needed) maintains its own state and represents a different stage in the processing pipeline. Typical tasks include the aggregation or filtering of inputs, the analysis of data, and the combination of the outputs of multiple predecessor nodes.

Between nodes, information is forwarded in the form of data tuples~(i.e., collections of key-value pairs) that only flow in one direction: from upstream nodes to downstream nodes. Communication with the outside world is usually handled via message queues~(see Figure~\ref{fig:data-stream-processing}), for example provided by systems such as Kafka~\cite{kreps11kafka} or Kestrel~\cite{kestrel}. If necessary, as it is the case for iterative computations~\cite{carbone15apache}, these queues can also be used to reinsert processed data into the application by relying \linebreak on the same queue as both output queue and input queue.

From a system operator's perspective, a stream processing framework offers several benefits with regard to executing applications. In the context of this paper, especially three aspects are of major importance: (1)~\textbf{Deployment \& Communication.} Provided with a logical graph of an application, the framework takes care of essential deployment tasks such as selecting servers, starting processing nodes, and setting up network connections. During execution, the framework then is responsible for serializing data tuples and routing them to their intended receiver nodes. In addition, frameworks commonly also comprise mechanisms for retransmitting tuples to tolerate network problems and for applying flow control to prevent nodes from being overwhelmed. (2)~\textbf{Scalability.} To improve performance by introducing parallelism, the framework allows to partition the workload and execute multiple instances of a node, as illustrated by the black boxes in Figure~\ref{fig:data-stream-processing}. This way, a stream processing application is able to exploit multiple cores and servers even within the same processing stage. (3)~\textbf{Automated Recovery.} Once an application is running, the framework continuously monitors the system for faults. In case it detects that a node instance has crashed, the framework automatically starts a new instance (possibly on a different server) and updates the routing configuration to ensure that the new instance is supplied with data.


\subsection{Problem Statement}

Creating a replication-protocol implementation that is ready for use in practice is a difficult and time-consuming task since many deployment, scalability, and recovery aspects need to be considered that are usually not part of the theoretical protocol specification~\cite{chandra07paxos}. The current version of the widely used replication library BFT-SMaRt~\cite{bessani14state}, for example, is the result of almost a decade of development, and even at this point it still does not support features such as the automated recovery of replicas. Our goal in this paper is to simplify the implementation of (existing and future) replication protocols by designing them as stream processing applications and thereby offloading most of the deployment and management tasks to the underlying framework. To be able to do so, we do not simply have to solve an engineering problem, but instead find a way to model replication protocols in the form of directed acyclic graphs, as required by stream processing frameworks. This is complicated by the fact that sub-protocols for consensus, garbage collection, and view changes are typically based on multiple phases of message exchange among the same group of replicas~(see Section~\ref{sec:background-replication}) and heavily draw on the existing circular dependencies. As detailed in the following section, our solution to this problem is to abandon the replica-oriented perspective commonly applied in existing protocol specifications and move to a data-oriented perspective that primarily concentrates on how information needs to flow through the system. That is, instead of relying on a few large replicas with complex states, we model a replication protocol as a graph of small processing nodes with comparably simple states that each perform a different task.


\section{\system}
\label{sec:approach}

This section presents \system, a replication protocol that has been tailored to run as an application on top of stream processing frameworks. In addition to the basic architecture, the section provides details on how \system handles requests, garbage-collects consensus information, and performs view changes. In the following, we focus on giving an intuition of \system's core concepts. For the full protocol specification please refer to the appendix of this paper.

\subsection{Overview}
\label{sec:approach-overview}

\begin{figure}
	\begin{center}
		\includegraphics{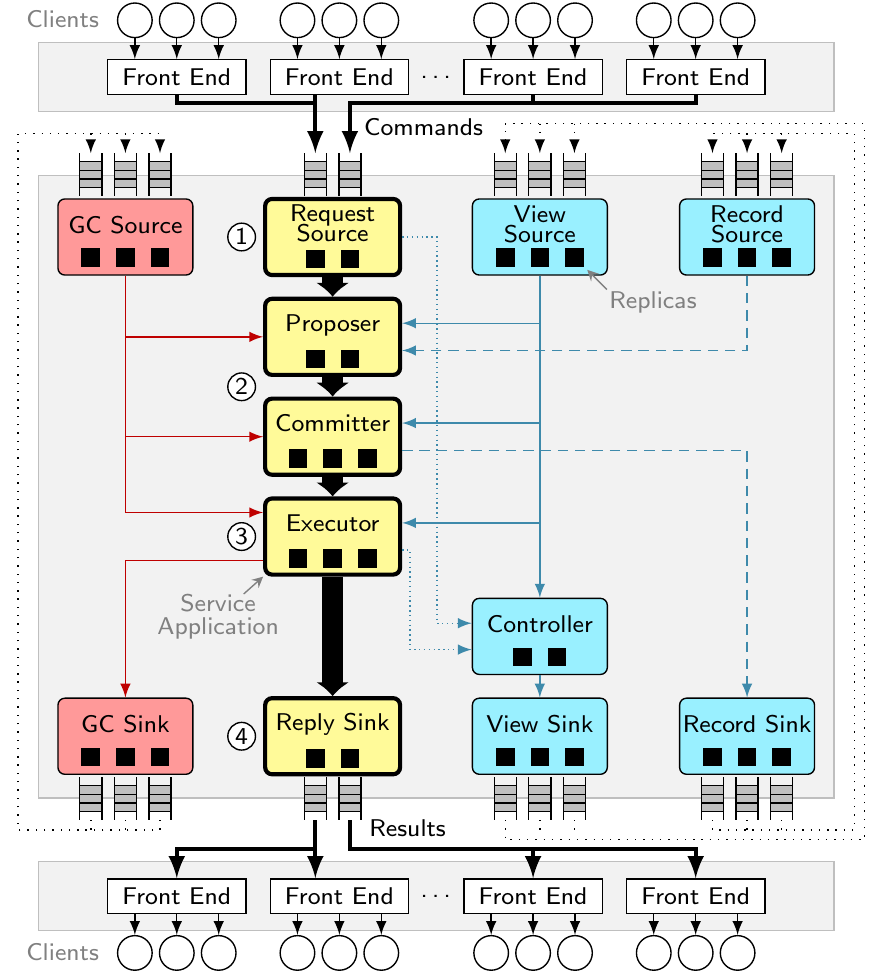}
	\end{center}
	\vspace{-3mm}
	\caption{Overview of the \system replication protocol, comprising nodes for request handling~(yellow), garbage collection~(red), and view changes~(blue).}
	\label{fig:architecture}
	\vspace{-.5mm}
\end{figure}

As illustrated in Figure~\ref{fig:architecture}, to meet the requirements of a stream processing application, the core of the \system replication protocol is structured as a directed acyclic graph in which information is forwarded as streams of data tuples. Based on their responsibilities, the nodes in the graph can be divided into three main categories: request handling~(yellow, see Section~\ref{sec:approach-request}), garbage collection~(red, see Section~\ref{sec:approach-gc}), and view change~(blue, see Section~\ref{sec:approach-view-change}). The replicated service application is integrated with the \emph{executor} stage.

For crash tolerance, each \system stage consists of multiple node instances, which in the following we refer to as replicas. Replicas belonging to the same stage are placed on separate servers, whereas replicas of different stages may be co-located. The number of replicas required per stage in Figure~\ref{fig:architecture} is symbolized by the number of black boxes. Two boxes indicate that a stage comprises $f+1$~replicas to tolerate up to $f$~replica crashes within the stage; three boxes represent $2f+1$~replicas.

The numbers in Figure~\ref{fig:architecture} mark \system's main workflow which consists of \numlabel{1}~receiving client commands that arrive through request input queues, \numlabel{2}~\,ordering these commands using a consensus algorithm, \numlabel{3}~\,executing the commands in the service application, and \numlabel{4}~placing the corresponding results in reply output queues. Clients with access to the queues are able to directly submit their commands to them and collect the results right away. All other clients typically communicate with \system through \emph{front-end} components that represent server-side proxies and act on the clients' behalf. If a result to a submitted command does not arrive within a configured amount of time~(e.g.,~due to the contacted front end having crashed), a client retries the operation by sending the command to another front end. As detailed in Section~\ref{sec:approach-request}, we designed \system to deal with command retransmissions by enabling executor replicas to detect and filter out duplicates.

In contrast to request handling, \system's mechanisms for garbage collection and view changes do not require external inputs, which is why their output queues are directly connected with their respective input queues to reinsert the emitted tuples back into the protocol~(see Figure~\ref{fig:architecture}). Independent of their specific type, all queues in \system are associated with a dedicated source or sink replica and therefore themselves do not have to provide any fault-tolerance guarantees. Specifically, we assume that a queue can crash as the result of a failure of its associated replica, and vice versa. Apart from using advanced queues such as Kafka~\cite{kreps11kafka}, this assumption for example makes it also possible to implement a queue as a simple Web service that is integrated with its replica, as done in our prototype.


\subsection{System Model}
\label{sec:approach-model}

\system makes only weak assumptions about the underlying stream processing framework and thus is compatible with a variety of existing systems~(e.g.,~\cite{toshniwal14storm,kulkarni15twitter,carbone15apache}). Specifically, it is not necessary for the framework to already provide consensus-based fault tolerance at a lower layer. At system startup, the framework needs to deploy all \system replicas and ensure that the replicas of neighboring stages know each other. Once the system is running, the primary task of the framework is to handle the communication between replicas by routing the emitted tuples to their designated receivers. \system allows the exchange of tuples to be performed using an asynchronous network but due to the FLP impossibility~\cite{fischer85impossibility} needs a partially synchronous environment to guarantee eventual progress, as it is the case for other replication protocols~\cite{lamport98part,ongaro14search,bessani14state}. \system clients and replicas (where necessary) are equipped with retransmission mechanisms that are responsible for ensuring eventual delivery even if the framework itself does not offer reliable transmission of tuples. However, as discussed in Section~\ref{sec:implementation}, \system can exploit built-in features to improve efficiency in case a framework provides stronger semantics.

\system tolerates up to $f$~simultaneous replica crashes in each protocol stage. If the underlying framework comprises means to automatically recover replicas and their states after a failure, the protocol is able to reintegrate them and thereby self-heal.

\subsection{Request Handling}
\label{sec:approach-request}

\system relies on a Paxos-style consensus algorithm to ensure that its executors process client commands in the same order and thereby remain consistent. The algorithm is view based and comprises three phases. First, a \emph{proposer} replica assigns monotonically increasing sequence numbers to newly incoming requests. In a second step, the requests are replicated to a set of \emph{committers} that store and confirm the sequence-number assignments. Finally, executors apply the commands in the determined order and forward their results. For this algorithm to be safe, only one of the proposer replicas at a time may make new proposals. The active proposer~$p$ is determined by the current view~$v$~(e.g.,~$p=v\,\%\,(f+1)$) and therefore loses its role as the result of a view change~(see Section~\ref{sec:approach-view-change}).

\headline{Consensus Process}
%
To invoke an operation~$op$ at the service application replicated by \system, a client~$c$ creates a \msg{Command}{c, t, op} and then~(either directly or via a front end) inserts the command into a request input queue; $t$ is a logical or physical timestamp selected to be greater than all timestamps of previous commands issued by this client. In combination, client ID~$c$ and timestamp~$t$ enable \system to uniquely identify a command~(e.g., to filter out duplicates, see below). When request-source replica~$r$ removes the command~$cmd$ from its input queue, the replica wraps it in a \msg{Request}{r, q, cmd} tuple and sends the request to the proposer stage; $q$ is a monotonically increasing request number that enables \system to track the request-handling progress.

Whenever the active proposer~$p$ of the current view~$v$ receives a request~$req$, it assigns a new sequence number~$s$ to the request and multicasts a \msg{Propose}{p, s, v, req} tuple to all committers. A committer~$k$ only accepts the proposal if it is also currently in view~$v$. In such case, the committer locally stores the sequence-number assignment for the request and attests this step in a \msg{Commit}{k, s, v, req} tuple to all executors. An executor accepts a request as soon as it has obtained \mbox{$f+1$}~commits from different committers for the same sequence number, provided that the commits have all been issued for the view the executor is currently in. At this point, the \linebreak request is committed and the consensus process complete.

\headline{Command Execution}
%
Executors perform duplicate detection to prevent the same command from being processed more than once. For this purpose, they maintain a vector~$\vec{T}_{exec}$ in which for each client they store the highest executed command timestamp. Furthermore, executors comprise a cache with the latest results to be able to respond to duplicate commands without having to reexecute them. Relying on these data structures, an executor processes committed requests in the order of their sequence numbers by performing the following steps. First, the executor extracts the client command from the committed request. Next, it compares the command's timestamp~$t_{cmd}$ to the client's locally stored timestamp~$t_{exec}$. If $t_{cmd}\leq{}t_{exec}$, the executor retrieves the result from the cache. Otherwise, the executor updates $t_{exec}$, invokes the command's operation, and caches the computed result. Finally, the executor sends the result (via reply sinks and possibly front end) to the client.

\subsection{Garbage Collection}
\label{sec:approach-gc}

To prevent replicas from running out of memory, each protocol stage in \system only maintains state for a limited amount of consensus instances, represented by a fixed-size window of sequence numbers whose lower bound is defined by a stability threshold~$s_{stable}$. If its window is full, a replica temporarily suspends its participation in the consensus process until it learns that $s_{stable}$ has increased. Raising the stability threshold is the main responsibility of \system's garbage-collection mechanism and triggered by periodic executor checkpoints of the application state. The rationale behind this approach is that once the effects of an executed command are reflected by a checkpoint, there is no longer a need to store consensus information about the command. As a result, replicas in such case are allowed to move their windows forward and garbage-collect information from all previous consensus instances.

\headline{Creating Checkpoints}
%
Executors in \system periodically create a checkpoint before processing a command with consensus sequence number~$s\,\%\,C\hspace{-.5mm}P=0$; $C\hspace{-.5mm}P$ is a configurable system-wide constant that represents the checkpoint interval. A checkpoint includes all essential information that is necessary to recreate the executor's state at this sequence number, which includes a snapshot of the service application, the vector~$\vec{T}_{exec}$ that is used for duplicate detection~(see Section~\ref{sec:approach-request}), as well as the result cache. Once the checkpoint is complete, the executor stores it at a predefined location. Stream processing frameworks such as Heron~\cite{kulkarni15twitter} for these purposes typically offer nodes access to local and remote file systems, and even cloud-based storage services. Having stored the checkpoint, the executor~$e$ in a last step sends sequence number~$s$ in a \msg{Checkpoint}{e, s} tuple to \system's garbage-collection sinks from where the tuple is forwarded to the garbage-collection sources via message queues~(see Figure~\ref{fig:architecture}).

\headline{Updating the Stability Threshold}
%
Garbage-collection source replicas maintain a vector~$\vec{S}_{cp}$ that for each executor contains the highest checkpoint sequence number learned from checkpoint notifications. They select the stability threshold~$s_{stable}$ to be the $f+1$ highest element in~$\vec{S}_{cp}$ as this guarantees that at least one copy of the corresponding checkpoint remains available, even if up to $f$~copies are no longer accessible~(e.g.,~due to storage-server crashes). Whenever the stability threshold increases, a source replica~$g$ emits a \msg{Stable}{g, s_{stable}} tuple \linebreak that is sent to all proposer, committer, and executor replicas.

Non-source replicas select the stability threshold as the $f+1$ highest value provided by different sources, which ensures that other replicas will eventually learn the same threshold, even if up to $f$~sources crash in the meantime. Each time the stability threshold increases, a replica adjusts its local consensus window accordingly and discards all information pertaining to lower sequence numbers. If an executor is lagging behind, for example as a result of asynchrony in the network, a window shift may cause the executor to skip sequence numbers. In order to catch up, the executor in such case first loads another executor's checkpoint for the stability threshold before continuing to process further commands.

\subsection{View Change}
\label{sec:approach-view-change}

\system's view-change mechanism enables the protocol to switch to another proposer in case the previously active proposer crashes. Decisions about whether a view change is needed or not are made by a set of \emph{controller} replicas that continuously monitor the progress of the consensus process. If the consensus gets stuck, the controllers announce a new view, which is then installed by the replicas of other stages.

\headline{Triggering a View Change}
%
The crash of the active proposer temporarily results in no new requests being proposed for ordering. Controllers in \system are responsible for detecting such a scenario and for this purpose constantly compare (1)~the number of incoming commands known to request sources with (2)~the number of commands whose consensus process report the executors as complete. To provide the controllers with the necessary information, each request source~$r$ periodically emits a \msg{Target}{r, q} tuple to all controllers, which includes the highest request number~$q$ the source has assigned to any request~(see Section~\ref{sec:approach-request}). In a similar way, each executor~$e$ periodically reports the consensus progress by emitting an \msg{Actual}{e, \vec{q}} tuple; $\vec{q}$ is a vector that for each request source contains the highest agreed request number. Relying on request numbers in the described way has the key benefit of enabling controllers to determine whether a request source has outstanding commands that so far have not been executed.

A controller triggers a view change if the number of completed requests does not increase on at least $f+1$~executors for a configurable amount of time, even though the corresponding source has reported the existence of new commands. In such case, the controller~$x$ increments a local view counter and announces the new view~$v$ in a \msg{View}{x, v} tuple. From this point on, the system-wide publication of the new view via view sinks and sources follows the same principle as the distribution of the stability threshold, which was detailed in Section~\ref{sec:approach-gc}.

\headline{Entering a New View}
%
To ensure that the outcomes of already completed consensus instances remain stable across a view change, the newly appointed active proposer must learn about the requests that might have committed in previous views, and therefore possibly were processed by at least one executor. \system solves this problem by requiring committers to maintain a record~$d=(s, v, req)$ for each sequence number~$s$ in their window; the record contains the most recent view~$v$ for which a committer has received a proposal as well as the associated request~$req$. Whenever a committer~$k$ learns about a higher view, it combines all of its records in a set~$\mathcal{D}$ and emits a \msg{Record}{k, \mathcal{D}} tuple to \system's record sinks. The sinks rely on their output queues to forward the tuples to the record \linebreak sources and finally to the active proposer of the new view.

Having been notified about a new view, the new proposer waits until it has obtained $f+1$~record tuples from different committers. This guarantees that if a request had previously passed the consensus process~(which requires confirmations from $f+1$~committers, see Section~\ref{sec:approach-request}), at least one of the received record tuples must contain the sequence-number assignment for the request. For each sequence number included in record tuples, the new proposer selects the associated request with the highest view as new proposal. After this procedure is complete, the proposer is allowed to suggest new requests \linebreak for all sequence numbers not covered by the record tuples.


\section{Parallelizing \system}
\label{sec:parallelization}

This section discusses how to increase scalability in \system by applying the concept of consensus-oriented parallelization~\cite{behl14scalable}. The main idea of this approach is to (1)~first distribute the responsibility for performing consensus across multiple partitions and then (2)~afterwards deterministically merge the outcomes of different partitions into a single sequence of commands. Thanks to the underlying stream processing framework handling tasks such as deployment and communication, the integration of partitioning into \system requires only minor modifications at the protocol level.


\begin{figure}[b!]
	\includegraphics{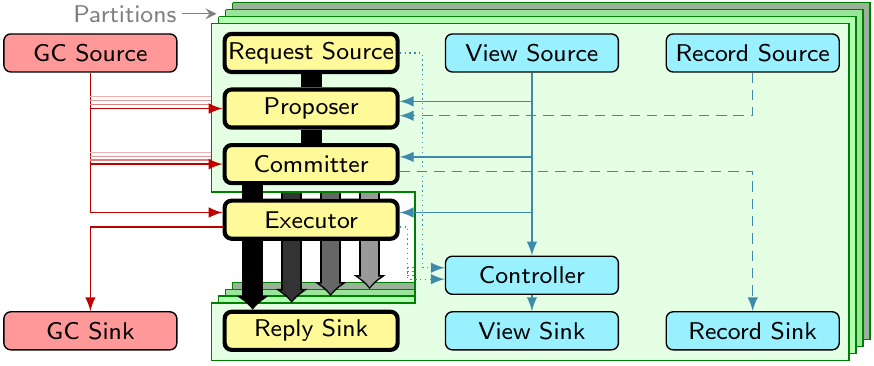}
	\vspace{-6mm}
	\caption{Use of partitions to parallelize consensus in \system.}
	\label{fig:partitions}
\end{figure}

As shown in Figure~\ref{fig:partitions}, each partition comprises its own sets of proposer and committer replicas in order to be able to run the agreement process independently of other partitions. Since proposers are partition local, so are the view-change stages responsible for switching to another proposer replica in case the old one crashes. In contrast, to minimize overhead we enable \linebreak all partitions to share the same garbage-collection replicas.

To ensure that all executors process requests in a consistent manner, they use a round-robin algorithm to compute a deterministic execution sequence number~$s_{exec}=s_i*P+i, \linebreak s_i\,\in\,\mathbb{N}, i\,\in\,=\{0,...,P-1\}$, with $s_i$ being the partition-local sequence number assigned to a command by a partition~$i$ and $P$ denoting the number of partitions. Once a command has been processed in the application, an executor forwards the result to the reply sinks of the corresponding partition.


\section{Implementation}
\label{sec:implementation}

Our \system prototype implementation\footnote{The prototype is publicly available under \url{https://gitlab.cs.fau.de/i4refit/tara}} is based on Heron~\cite{kulkarni15twitter}, a stream processing framework that, for example, is currently used in production at Twitter.
\system offers similar client and application interfaces as existing replication libraries. A client or front end can asynchronously submit a request to any request source and will receive a result over the same channel.
Since requests and results are represented as simple \linebreak byte strings, both can have an arbitrary size and structure.

As typical for Heron, each \system node~(see Figure~\ref{fig:architecture}) runs in an independent process with its own Java virtual machine. This strongly improves the fault isolation in the system, as the failure of one node does not affect others, even when they run on the same server. It additionally allows for easy debugging, because each node can be accessed and analyzed individually.

In general, \system makes it possible to distribute its nodes across a large number of servers. However, for better comparability with traditional replication libraries we co-locate replicas of different stages~(e.g.,~a proposer, a committer, an executor) on the same server. Furthermore, for improved efficiency we integrate the reply sinks with their corresponding executors.

As discussed in Section~\ref{sec:approach-model}, the \system protocol where necessary specifies its own retransmission mechanisms in order to ensure progress in the presence of an unreliable network. Implementing \system on top of Heron, we are able to outsource some of the retransmission logic to the underlying framework by exploiting Heron's built-in support for at-least-once delivery of tuples that originate in source nodes.


\section{Evaluation}
\label{sec:evaluation}

This section evaluates our prototype implementation of \system, while relying on the widely used replication library BFT-SMaRt~\cite{bessani14state} as baseline. As the name suggests, BFT-SMaRt was originally designed to tolerate Byzantine faults, however for a fair comparison with \system we only evaluate the library in its later added configuration for crash tolerance.

\subsection{Development Effort}

In order to get an impression of how our approach simplifies the development of replication protocols, we analyze the size of \system's code base compared with BFT-SMaRt~(version~1.2).

\headline{Methodology}
%
To ensure meaningful results, we only consider functionality that is present in both implementations. For BFT-SMaRt, this for example means that we exclude all code that is solely required and executed if the system is configured to tolerate Byzantine faults. Specifically, our analysis focuses on the core parts of the two protocols, namely the mechanisms for request handling, garbage collection, and view change.

Both the BFT-SMaRt and \system implementation are written in Java and use a similar coding style, which allows us to perform the analysis based on the number of code lines spent on a specific functionality. Of course, examining code-line counts does not necessarily reveal all the complexity that might be contained in an implementation, however it gives us the big picture of what is necessary to develop a replication infrastructure. Since we are interested in the effort it takes a programmer to implement a protocol, we exclude code lines that are trivial~(e.g., due to only comprising a closing bracket), usually automatically inserted by a programmer's development environment~(e.g.,~import statements), or have no impact on the running system~(e.g.,~empty lines and comments).

\headline{Results}
%
Table~\ref{tab:evaluation-code} presents the findings of our analysis. Apart from the main mechanisms, we also report numbers for two additional categories: data structures that cannot necessarily be attributed to one specific mechanism, and code parts handling general infrastructure tasks such as system startup or the communication between replicas. Overall, the results show that thanks to leveraging Heron as underlying platform, \system's implementation is only about a third of the size of BFT-SMaRt.


\begin{table}[b!]
	\vspace{-.4mm}
	\caption{Code-base comparison between BFT-SMaRt and \system}
	\label{tab:evaluation-code}
	\vspace{-6mm}
	\begin{center}
		\renewcommand{\arraystretch}{1.05}
		\begin{tabular}{l|r|r|r}
			& \multicolumn{1}{c|}{\textbf{BFT-SMaRt}} & \multicolumn{1}{c|}{\textbf{\system}} & \multicolumn{1}{c}{\textbf{Difference}}\\
			\hline
			\hline
			Request Handling & 905 LoC & 274 LoC & --\,70\%\\
			\hline
			Garbage Collection & 551 LoC & 185 LoC & --\,66\%\\
			\hline
			View Change & 867 LoC & 297 LoC & --\,66\%\\
			\hline
			\hline
			Data Structures & 756 LoC & 421 LoC & --\,44\%\\
			\hline
			Infrastructure & 918 LoC & 299 LoC & --\,67\%\\
			\hline
			\hline
			\textbf{Total} (LoC: \underline{L}ines \underline{o}f \underline{C}ode) & 3,997 LoC & 1,476 LoC & --\,63\%\\
		\end{tabular}
	\end{center}
\end{table}


Specifically, our analysis enables us to make three key observations: (1)~Attributing code parts to one of the five categories was not always straightforward in case of BFT-SMaRt as the implementation often intertwines multiple mechanisms within the same class, occasionally even within the same method. In contrast, \system's design cleanly separates the different mechanisms by implementing them in dedicated nodes~(as illustrated in Figure~\ref{fig:architecture}). For nodes that need to participate in more than one mechanism~(e.g.,~executors), this separation is still visible within the node implementation in the form of different handlers for different types of incoming tuples~(e.g.,~commit tuples or garbage-collection tuples). Altogether, \system's architecture makes it not only easier to analyze the code, but also significantly simplifies development, maintenance, and debugging. (2)~A considerable amount of protocol-level code in BFT-SMaRt is spent on the synchronization of workflows that are implemented in different threads. \system, on the other hand, leaves most of the concurrency handling to the underlying Heron and consequently minimizes the need for programmers to deal with synchronization themselves, a task that is notoriously error-prone. \linebreak (3)~BFT-SMaRt itself comprises means for typical infrastructure duties such as establishing network connections, dispatching received messages, and handling communication failures, all of which are examples of functionality that in \system is provided by Heron. Therefore, most code lines in \system dedicated to infrastructure are used to configure the Heron platform and specify the node graph representing the protocol. As a consequence, in the infrastructure category \system allows a \linebreak code-size reduction of about~67\% compared with BFT-SMaRt.


\begin{figure}
	\begin{minipage}{.47\columnwidth}
		\vspace{-.5mm}
		\begin{center}
			\includegraphics[page=1]{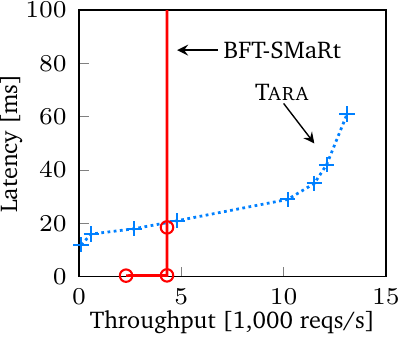}
		\end{center}
		\vspace{-5mm}
		\caption{Throughput versus latency results for non-batched consensus.}
		\label{fig:evaluation-performance-nobatch}
	\end{minipage}
	\hfill
	\begin{minipage}{.48\columnwidth}
		\begin{center}
			\includegraphics[page=2]{figures/evaluation-throughput.pdf}
		\end{center}
		\vspace{-5mm}
		\caption{Throughput versus latency for batched consensus (batch size: 5).}
		\label{fig:evaluation-performance-batch}
	\end{minipage}
	\vspace{-1.5mm}
\end{figure}

\subsection{Performance}

\vspace{-.5mm}

To evaluate the performance of \system in comparison with BFT-SMaRt, we conduct experiments with a coordination service~\cite{distler16resource} which we integrate with both systems. For the replicas we use a cluster of three servers~(Intel Xeon CPU E3-1275, 3.6\,GHz, 16\,GB RAM); the clients are hosted by up to five additional machines. We configure the clients to submit requests in a closed loop, meaning that each client waits with the transmission of a new request until it has received the result to its previous one. The performance results reported in the following represent the average of three runs each.

\headline{Non-batched Consensus}
%
In our first experiment, we configure BFT-SMaRt and \system to order one client request per consensus instance, since this setting is a stress test for the request-handling mechanism and hence ideal for assessing the efficiency of the replication protocol. As shown in Figure~\ref{fig:evaluation-performance-nobatch}, when we increase the workload BFT-SMaRt quickly reaches saturation at a throughput of about 4,300 requests per second. This behavior is caused by the fact that BFT-SMaRt's consensus sub-protocol strictly sequentializes the agreement process by executing at most one consensus instance at a time. Consequently, once the system reaches a point at which there is always a consensus instance active, it is no longer able to further increase throughput. This bottleneck in turn causes the latency to increase considerably, as also visible in Figure~\ref{fig:evaluation-performance-nobatch}.

Compared with BFT-SMaRt \system starts with a higher la\-ten\-cy, which is mainly a result of the inter-process communication overhead in Heron. However, in contrast to BFT-SMaRt, \system is able to keep response times low at higher throughputs due to processing consensus instances in a pipelined manner. That is, as long as there are free slots in the sequence-num\-ber window~(see Section~\ref{sec:approach-gc}), the proposer in \system can submit new proposals without the need to wait for the previous consensus instances to complete. This pipelining allows \system to handle up to 13,000 requests per second in this setting, which is about three times the throughput of BFT-SMaRt.

\headline{Batched Consensus}
%
To reduce the agreement overhead per request, replication protocols commonly offer the possibility to order a batch of multiple requests within the same consensus instance~\cite{bessani14state}. For our second experiment, we implement this technique in \system by enabling request sources to combine incoming client commands and forward them as a single batch request. Figure~\ref{fig:evaluation-performance-batch} presents the results of this experiment with a maximum batch size of 5 in comparison with BFT-SMaRt. In essence, the performance numbers show the same picture as in the non-batched consensus case, only at higher absolute throughput of close to 40,000~requests per second.
This clearly shows that \system can benefit from batching.

Besides batching, other common protocol-level optimizations such as read optimization or tentative execution could be integrated into \system as well. On the other hand, optimizations relying on a monolithic system layout (e.g., fast paths between steps on the same replica) are not applicable to \system as each step by design resides in its own node.


\begin{figure}
	\begin{minipage}{.405\columnwidth}
		\begin{center}
			\includegraphics{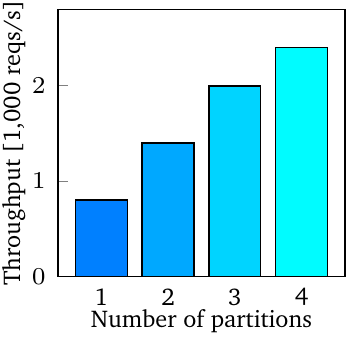}
		\end{center}
		\vspace{-5mm}
		\caption{\system throughput for different numbers of partitions.}
		\label{fig:evaluation-partitions}
	\end{minipage}
	\hfill
	\begin{minipage}{.54\columnwidth}
		\begin{center}
			\includegraphics{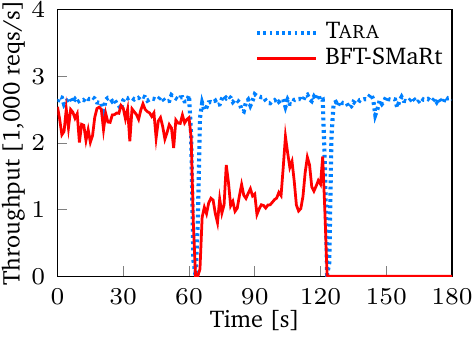}
		\end{center}
		\vspace{-5mm}
		\caption{Impact of proposer failures on the throughput of BFT-SMaRt and \system.}
		\label{fig:evaluation-faults}
	\end{minipage}
	\vspace{-1.5mm}
\end{figure}

\subsection{Partitions}

Parallelizing consensus, as done by the partitioned variant of \system~(see Section~\ref{sec:parallelization}), is especially relevant in use-case scenarios in which the agreement process constitutes the performance bottleneck of the system~\cite{behl15consensus}, for example due to the nodes involved being resource constrained. In our third experiment, we create such a setting by individually throttling the proposer so that overall throughput is now at about 800~requests per second. As shown in Figure~\ref{fig:evaluation-partitions}, in such a setting the use of multiple partitions enables \system to scale by parallelizing consensus across a larger number of (also resource-constrained) nodes. Notice that BFT-SMaRt does not support partitioned consensus, which is why for this experiment we only report measurement results for \system.

\subsection{Fault Tolerance}

In our fourth experiment, we expose the two systems to leader failures in order to evaluate their view-change mechanisms. As shown in Figure~\ref{fig:evaluation-faults}, when we deliberately crash the current leader replica~(i.e.,~the active proposer node in \system) after 60~seconds of uninterrupted service, both systems trigger a view change to assign the leader role to a different replica. While without leader, during the view change no new requests can be agreed on and consequently the throughput temporarily drops to zero before eventually returning to a higher level. When we repeat the procedure with the newly elected leader 120~seconds into the experiment, BFT-SMaRt becomes unavailable since a single remaining replica is insufficient to safely make progress. In contrast, \system is able to continue request processing even after the second proposer crash, because Heron in the meantime has automatically restarted the first proposer replica after having detected its failure. This example scenario illustrates a key benefit of implementing replication protocols on top of stream processing frameworks such as Heron: \system can leverage Heron's support for automated recovery, a feature that is commonly not part of traditional replication libraries.


\section{Related Work}
\label{sec:related}

Replication protocols are notoriously difficult to implement, which is why several previous works have aimed at easing their development. For this reason, for example, a large body of secondary literature exists whose main purpose is to give advice on how to implement replication protocols such as Paxos based on their specification~(e.g.,~\cite{chandra07paxos,kirsch08paxos,distler21byzantine}). Ongaro~et~al.~\cite{ongaro14search} even went one step further by designing a replication protocol from scratch and targeting understandability as most important property. In this paper, we have shown stream-based replication to be an approach that effectively minimizes the number of problems a programmer needs to worry about when implementing a replication protocol. Still, the resulting system is able to match (or in some cases even exceed) the efficiency of \linebreak traditional replication libraries, as confirmed by our evaluation.

Replication is an essential concept when it comes to providing fault tolerance in stream processing systems~\cite{balazinska05fault,martin11low,liu17storm,martin19low}, however notice that previous works in this area had a fundamentally different focus than \system. While other researchers aimed at providing replication-based fault tolerance to stream processing applications, \system itself is a stream processing application that offers fault tolerance to arbitrary network-based services. That is, instead of integrating replication mechanisms into the framework and tailoring them to the specific characteristics of stream processing applications~\cite{balazinska05fault,martin11low,liu17storm,martin19low}, \system implements state-machine replication on top of such a framework in a generic manner.

With stream processing systems playing an important role in production, over the years significant efforts have been made to improve different aspects of these frameworks. Among other things, this includes approaches to optimize the deployment of processing nodes on the machines available~\cite{aniello13adaptive}, techniques to minimize energy consumption based on the current workload~\cite{eibel18strome}, and mechanisms to retain high quality-of-service levels even in the presence of load spikes~\mbox{\cite{tatbul07staying,rivetti20load}}. Since \system itself is a stream processing application, it is able to benefit from many existing (and potentially future) improvements to underlying stream processing frameworks. For example, apart from the fact that replicas of the same stage must not be executed on the same server, \system imposes no restrictions on the mapping of processing nodes to machines, and \linebreak consequently can profit from optimized placement algorithms.


\section{Conclusion}
\label{sec:conclusion}

In this paper we presented stream-based state-machine replication, an approach that leverages stream processing frameworks to significantly simplify the development, deployment, and operation of general-purpose replication protocols. The analysis of our first stream-based protocol \system shows that our method reduces code size by about two thirds compared with the BFT-SMaRt library. Even though the additional layer of the stream processing engine leads to an increase in latency, \system is still able to sustain throughputs of tens of thousands of requests per second. Additionally, the automated recovery mechanism of the stream processing engine allows \system to automatically restart and reintegrate failed leaders without any necessary manual actions.


\vfill

\noindent{\small{}\emph{Acknowledgments}: This work was partially supported by the German\vspace{-.7mm} Research Council (DFG) under grant no. DI 2097/1-2~(``REFIT'').}

\bibliographystyle{IEEEtran}
\bibliography{bib/paper-tiny}

\appendices
\section{\system Protocol Specification}

This specification describes \system, a crash-fault tolerant state-machine replication protocol designed to run on top of a stream processing framework such as Apache Heron.
The specification includes all components that are part of the stream processing workflow of \system, it does not encompass components outside of the core system such as clients and incoming/outgoing queues.

The specification first introduces all required data structures, grouped by their designated task in the system. Afterwards, the protocol nodes of \system are specified. They can be split into three groups: sources, sinks and processing nodes.

Sources represent the start of the stream processing workflow and generate tuples based on an input queue that is fed by an outside source (i.e., clients/frontends or queues from the corresponding sinks).
Sinks, on the other hand, represent the end of the stream processing workflow. Reply sinks pass responses to the client whereas the other sinks are responsible for feeding garbage collection information back to their corresponding sources (e.g., by using a shared queue).
As all sinks work in the same fashion, they are not specified individually but are all represented by a common \textsc{SINK} class.
All nodes in between a source and a sink are processing nodes: They receive tuples from preceding nodes, process those, and forward the resulting tuples to subsequent nodes.

The nodes in \system have the following tasks:

\begin{description}[leftmargin=7.8em,style=nextline]
	\item[Request Source:] Receives requests from clients (or frontends) and passes them into the consensus algorithm.
	\item[GC Source:] Forwards garbage collection information to processing nodes.
	\item[View Source:] Forwards information about the current view to processing nodes.
	\item[Record Source:] Forwards records to processing nodes when a view change was issued.
	\item[Proposer:] In each view, one acts as the current leader and assigns a sequence number to each issued request.
	\item[Committer:] Acknowledges and forwards the sequence number/request-tuples proposed by the current proposer.
	\item[Executer:] Executes requests after enough committers acknowledged them and previous requests have been executed.
	\item[Controller:] Monitors whether the current proposer is active and issues a view change if not.
	\item[Request Sink:] Sends responses back to clients (or frontends).
	\item[GC Sink:] Feeds the current garbage collection information back to the GC sources.
	\item[View Sink:] Feeds information on the current view back to the view sources.
	\item[Record Sink:] Feeds records to the record sources when a view change was issued.
\end{description}

To tolerate up to \textit{f} crash faults, the system must contain the number of nodes as shown in Table \ref{tab:nodes}.

\begin{table}[t]
	\renewcommand{\arraystretch}{1.2}
	\centering
	\caption{Number of nodes required to tolerate up to \textit{f} faults}
	\label{tab:nodes}
	\begin{tabular}{l|r}
		\textbf{Node Type} & \textbf{\# of Nodes} \\\hline
		Request Source & $f+1$ \\
		GC Source & $2f+1$ \\
		View Source & $2f+1$ \\
		Record Source & $2f+1$ \\
		& \\
		Proposer & $f+1$ \\
		Committer & $2f+1$ \\
		Executor & $2f+1$ \\
		Controller & $f+1$ \\
		& \\
		Request Sink & $f+1$ \\
		GC Sink & $2f+1$ \\
		View Sink & $2f+1$ \\
		Record Sink & $2f+1$
	\end{tabular}
\end{table}

Furthermore, the specification makes some assumptions on the interaction of \system with the stream processing framework underneath:
For sources, the \texttt{produce()}-function specifies how the next tuple of a source is generated.
It is assumed that the stream processing framework offers an interface that lets sources emit tuples on a regular basis, as is done by the \texttt{nextTuple()}-method of the \texttt{ISpout} interface in Heron.
For sinks and processing nodes, the stream processing framework must offer a way to receive tuples from preceding nodes. In the specification, it is assumed these are passed to the \texttt{process()}-methods that each processing node/sink owns. Additionally, sources and processing nodes assume the stream processing framework offers an \texttt{emit()}-method to pass a tuple to a specified group of subsequent nodes.
Loops of the form ``For each \textsc{Type} \texttt{t} in [\texttt{a}, \texttt{b}]'' define the start index \texttt{a} as inclusive and the end index \texttt{b} as exclusive.

\clearpage


\section*{Supplementary material}

# Data Structures

## Helper Functions

```
1 /* If−then−else */
2 ANY ite(BOOLEAN v; ANY a, ANY b) {
3   If (v == true) return a;
4   return b;
5 }
```

## Basic Data Structures

```
1 interface SET<V> {
2   /* Operations */
3   void add(V value);
4 }

5 interface MAP<K, V> {
6   /* State */
7   SET<K> keys;
8   SET<V> values;
9
10  /* Operations */
11  VOID put(K k, V v);              /* Key accessed via [] operator */
12  V get(K k);                      /* Accessed via [] operator */
13 }
```

## Window

```
1 class WINDOW<V> {
2   /* State */
3   NUMBER capacity;
4   NUMBER min;
5   NUMBER max;
6   NUMBER pos;
7   V[] values;
8
9   /* Constructor */
10  WINDOW(NUMBER min, NUMBER max) {
11    capacity := max − min;
12    min := min;
13    max := max;
14    pos := min;
15    values := V[capacity];
16  }
17
18  /* Operation */
19  VOID put(NUMBER index, V value) {     /* Index accessed via [] operator */
20    /* Check state and input */
21    If (pos == max) return;
22    If (pos != index) return;
23
24    /* Update state */
25    values[index − min] := value;
26    pos := index + 1;
27  }
28
29  V get(NUMBER index) {                 /* Index accessed via [] operator */
30    If (index < min) return nil;
31    If (index >= pos) return nil;
32    return values[index − min];
33  }
34
35  /* Operations */
36  VOID fill(NUMBER to) {
37    If (to <= pos) return;
38    If (to > max) return;
39    pos = to;
40  }
41
42  VOID move(NUMBER min) {
43    /* Only move forward */
44    If (min <= min) return;
45
46    /* Determine state */
47    V[] vs := V[capacity];
48    For each NUMBER index in
          [min, min + capacity]: vs[index − min] := this[index];
49
50    /* Update state */
51    min := min;
52    max := min + capacity;
53    pos := max(pos, min);
54    values := vs;
55  }
56
57  VOID clear(NUMBER from) {
58    NUMBER start := max(from, min);
59    For each NUMBER index in [start, pos]: values[index − min] := nil;
60    pos := start;
61  }
62
63  VOID reset() {
64    clear(min);
65  }
66
67  BOOLEAN appendable(NUMBER index) {
68    If (pos == max) return false;
69    return (pos == index);
70  }
71
72  VOID append(V value) {
73    put(pos, value);
74  }
75 }
```

```
76 typedef WINDOWS<I is an ID, V>: MAP<I, WINDOW<V>>
```

## Comparison

```
1  class NUMBEROPINIONS<I is an ID> extends MAP<I, NUMBER> {
2     /* Constructor */
3     NUMBEROPINIONS() {
4        For each I id: this[id] := 0;
5     }
6
7     /* Operation */
8     NUMBER highest(NUMBER threshold) {
9        NUMBER[] ranking := values sorted in descending order;
10       return ranking[threshold − 1];
11    }
12 }

13 class PROGRESSOPINIONS<I is an ID, P is a MAP<D is an ID, NUMBER>>
                extends MAP<I, P> {
14    /* Constructor */
15    PROGRESSOPINIONS() {
16       For each I key {
17          this[key] := MAP();
18          For each D id: this[key][id] := 0;
19       }
20    }
21
22    /* Operation */
23    P highest(NUMBER threshold) {
24       P result := MAP();
25       For each D id {
26          NUMBEROPINIONS opinions := NUMBEROPINIONS<I>();
27          For each I key: opinions[key] := this[key][id];
28          result[id] := opinions.highest(threshold);
29       }
30       return result;
31    }
32 }

33 class WINDOWOPINIONS<I is an ID, V> extends MAP<I, WINDOW<V>> {
34    /* Constructor */
35    WINDOWOPINIONS(NUMBER min, NUMBER max) {
36       For each I id: this[id] := WINDOW(min, max);
37    }
38
39    /* Operations */
40    VOID fill(NUMBER to) {
41       For each I id: this[id].fill(to);
42    }
43
44    VOID move(NUMBER min) {
45       For each I id: this[id].move(min);
46    }
47
48    VOID sync(WINDOW<NUMBER, *> window) {
49       For each I id: this[id].sync(window);
50    }
51
52    VOID sync(N min) {
53       For each I id {
54          this[id].move(min);
55          this[id].clear(min);
56       }
57    }
58
59    NUMBER available(N index) {
60       NUMBER count := 0;
61       For each I id {
62          If (index < this[id].pos) count++;
63       }
64       return count;
65    }
66 }
```

## Observer

```
1  class OBSERVER<I is an ID> {
2     /* State */
3     NUMBEROPINIONS<I> thresholds;
4     NUMBER current;
5
6     /* Initialization */
7     OBSERVER() {
8        thresholds := NUMBEROPINIONS();
9        current := 0;
10    }
11
12    BOOLEAN update(I index, NUMBER threshold) {
13       /* Check input */
14       If (threshold <= current) return false;
15       If (threshold <= thresholds[index]) return false;
16
17       /* Store input */
18       thresholds[index] := threshold;
19
20       /* Update state */
21       NUMBER old := current;
22       current := thresholds.highest(F+1);
23       return (current != old);
24    }
25 }
```

## Protocol Data Structures

```
1  class PAYLOAD {
2     /* State */
3     CLIENTID xid;
4     COMMANDNR xnr;
5     ANY payload;
6
7     /* Constructor */
8     PAYLOAD(CLIENTID xid, COMMANDNR xnr, ANY payload) {
9        xid := xid;
10       xnr := xnr;
11       payload := payload;
12    }
13 }

14 class REQUEST {
15    /* State */
16    REQUESTSOURCEID source;
17    REQUESTNR rnr;
18    COMMAND command;
19
20    /* Constructor */
21    REQUEST(REQUESTSOURCEID source, REQUESTNR rnr,
               COMMAND command) {
22       source := source;
23       rnr := rnr;
24       command := command;
25    }
26 }

27 class SNAPSHOT {
28    /* State */
29    STATE state;
30    MAP<CLIENTID, COMMANDNR> filter;
31    MAP<CLIENTID, RESULT> results;
32
33    /* Constructor */
34    SNAPSHOT(STATE state, MAP<CLIENTID, COMMANDNR> filter,
               MAP<CLIENTID, RESULT> results) {
35       state := state;
36       filter := filter;
37       results := results;
38    }
39 }

40 class RECORD {
41    /* State */
42    VIEWNR view;
43    REQUEST request;
44
45    /* Initialization */
46    RECORD(VIEWNR view, REQUEST request) {
47       view := view;
48       request := request;
49    }
50 }

51 typedef COMMAND: PAYLOAD;
52 typedef RESULT: PAYLOAD;
53 typedef PROGRESS: MAP<REQUESTSOURCEID, REQUESTNR>;
```

## Tuples

```
1  class TUPLE {
2    /* State */
3    NODEID creator;
4
5    /* Initialization */
6    TUPLE(NODEID i) {
7      creator := i;
8    }
9  }

10 class REQUESTTUPLE extends TUPLE {
11   /* State */
12   REQUEST request;
13
14   /* Initialization */
15   REQUESTTUPLE(NODEID i, REQUEST r) : TUPLE(i) {
16     request := r;
17   }
18 }

19 class CONSENSUSTUPLE extends TUPLE {
20   /* State */
21   PARTITION partition;
22   SEQNR snr;
23   VIEWNR view;
24   REQUEST request;
25
26   /* Initialization */
27   CONSENSUSTUPLE(NODEID i, PARTITION p, SEQNR s, VIEWNR v,
                    REQUEST r) : TUPLE(i) {
28     partition := p;
29     snr := s;
30     view := v;
31     request := r;
32   }
33 }

34 class RESULTTUPLE extends TUPLE {
35   /* State */
36   RESULT result;
37
38   /* Initialization */
39   RESULTTUPLE(EXECUTORID i, RESULT r) : TUPLE(i) {
40     result := r;
41   }
42 }

43 class GCTUPLE extends TUPLE {
44   /* State */
45   CHECKPOINTNR cnr;
46   SET<EXECUTORID> executors;
47
48   /* Initialization */
49   GCTUPLE(NODEID i, CHECKPOINTNR c, SET<EXECUTORID> es) :
             TUPLE(i) {
50     cnr := c;
51     executors := es;
52   }
53
54   /* Operation */
55   SEQNR snr() {
56     return (cnr * CP_INTERVAL) / #PARTITIONS;
57   }
58 }

59 class PROGRESSTUPLE extends TUPLE {
60   /* State */
61   PROGRESS progress;
62
63   /* Initialization */
64   PROGRESSTUPLE(NODEID i, REQUESTNR r) : TUPLE(i) {
65     progresses := PROGRESS();
66     progresses[i] := r;
67   }
68
69   PROGRESSTUPLE(NODEID i, PROGRESS p) : TUPLE(i) {
70     progress := p;
71   }
72 }
```

```
73 class VIEWTUPLE extends TUPLE {
74   /* State */
75   PARTITION partition;
76   VIEWNR view;
77
78   /* Initialization */
79   VIEWTUPLE(NODEID i, PARTITION p, VIEWNR v) : TUPLE(i) {
80     partition := p;
81     view := v;
82   }
83 }

84 class RECORDTUPLE extends TUPLE {
85   /* State */
86   PARTITION partition;
87   VIEWNR view;
88   WINDOW<RECORD> records;
89
90   /* Initialization */
91   RECORDTUPLE(NODEID i, PARTITION p, VIEWNR v,
               WINDOW<RECORD> rs) : TUPLE(i) {
92     partition := p;
93     view := v;
94     records := rs;
95   }
96 }
```

# Protocol Nodes

## Nodes

```
1  class NODE {
2    /* State */
3    NODEID id;
4
5    /* Initialization */
6    TUPLE(NODEID i) {
7      id := i;
8    }
9  }

10 class SOURCE extends NODE {
11   /* Initialization */
12   SOURCE(NODEID i) : TUPLE(i) {}
13 }

14 class PROCESSOR extends NODE {
15   /* Initialization */
16   PROCESSOR(NODEID i) : TUPLE(i) {}
17
18   /* Output */
19   VOID emit(TUPLE t, SET<NODEID> is) {
20     If (t != nil)  Send tuple t to the nodes represented by the IDs in is;
21   }
22 }

23 class SINK extends NODE {
24   /* State */
25   QUEUE queue;                                    /* External queue */
26
27   /* Initialization */
28   SINK(NODEID i, QUEUE q) : TUPLE(i) {
29     queue := q;
30   }
31
32   /* Tuple handling */
33   VOID process(TUPLE t) {
34     queue.append(t);
35   }
36 }
```

## Request Source

```
1  class REQUESTSOURCE extends SOURCE {
2    /* State */
3    PARTITION partition;
4    QUEUE<COMMAND> commands;              /* External queue */
5    REQUESTNR next;
6
7    /* Initialization */
8    REQUESTSOURCE(REQUESTSOURCEID i, PARTITION p,
             QUEUE<COMMAND> q) : SOURCE(i) {
9      partition := p;
10     commands := q;
11     next := 0;
12   }
13
14   /* Command handling */
15   VOID produce() {
16     /* Create and emit request */
17     COMMAND c := commands.dequeue();
18     REQUEST r := REQUEST(id, next, c);
19     REQUESTTUPLE request := REQUESTTUPLE(id, r);
20     emit(request, PROPOSERS(partition));
21
22     /* Update state */
23     next := next + 1;
24   }
25
26   /* View change */
27   Periodically {
28     PROGRESSTUPLE progress := PROGRESSTUPLE(id, next);
29     emit(progress, CONTROLLERS(partition));
30   }
31 }
```

## GC Source

```
1  class GCSOURCE extends SOURCE {
2    /* State */
3    QUEUE<GCTUPLE>[] gcs;                 /* External queues */
4    GCTUPLE output;
5    OBSERVER<EXECUTORID> gc;
6
7    /* Initialization */
8    GCSOURCE(GCSOURCEID i, QUEUE<GCTUPLE>[] qs) : SOURCE(i) {
9      gcs := qs;
10     output := GCTUPLE(id, 0, nil);
11     gc := OBSERVER();
12   }
13
14   /* Garbage collection */
15   VOID produce() {
16     /* Update state */
17     GCTUPLE gt := gcs[id].dequeue();
18     SET<EXECUTORID> es := SET();
19     If (gt.creator is a GCSOURCEID) {
20       If (gt.cnr <= gc.current) return;
21       gc.current := gt.cnr;
22       es := gt.executors;
23     } else if (gt.creator is a GCSINKID) {
24       If (gc.update(gt.creator, gt.cnr) == false) return;
25       For each EXECUTORID e {
26         If (gc.thresholds[e] >= gc.current) es.add(e);
27       }
28     }
29
30     /* Publish new garbage-collection threshold */
31     output := GCTUPLE(id, gc.current, es);
32     emit(output, PROPOSERS());
33     emit(output, COMMITTERS());
34     emit(output, EXECUTORS());
35   }
36
37   Periodically {
38     For each GCSOURCEID i {
39       If (i != id) gcs[i].append(output);
40     }
41   }
42 }
```

## View Source

```
1  class VIEWSOURCE extends SOURCE {
2    /* State */
3    PARTITION partition;
4    VIEWTUPLE output;
5    QUEUE<VIEWTUPLE>[] views;             /* External queues */
6    OBSERVER<CONTROLLERID> view;
7
8    /* Initialization */
9    VIEWSOURCE(VIEWSOURCEID i, PARTITION p,
             QUEUE<VIEWTUPLE>[] qs) : SOURCE(i) {
10     partition := p;
11     views := qs;
12     output := VIEWTUPLE(id, partition, 0);
13     view := OBSERVER();
14   }
15
16   /* View change */
17   VOID produce() {
18     /* Update state */
19     VIEWTUPLE vt := views[id].dequeue();
20     If (vt.creator is a VIEWSOURCEID) {
21       If (vt.view <= view.current) return;
22       view.current := vt.view;
23     } else if (vt.creator is a VIEWSINKID) {
24       If (view.update(vt.creator, vt.view) == false) return;
25     }
26
27     /* Publish new view */
28     output := VIEWTUPLE(id, partition, view.current);
29     emit(output, PROPOSERS(partition));
30     emit(output, COMMITTERS(partition));
31     emit(output, CONTROLLERS(partition));
32     emit(output, EXECUTORS());
33   }
34
35   Periodically {
36     For each VIEWSOURCEID i {
37       If (i != id) views[i].append(output);
38     }
39   }
40 }
```

## Record Source

```
1  class RECORDSOURCE extends SOURCE {
2    /* State */
3    PARTITION partition;
4    QUEUE<RECORDTUPLE> records;           /* External queue */
5
6    /* Initialization */
7    RECORDSOURCE(RECORDSOURCEID i, PARTITION p,
             QUEUE<RECORDTUPLE> q) : SOURCE(i) {
8      partition := p;
9      records := q;
10   }
11
12   /* View change */
13   VOID produce() {
14     RECORDTUPLE record := records.dequeue();
15     emit(record, PROPOSERS(partition));
16   }
17 }
```

## Proposer

```
1  class PROPOSER extends PROCESSOR {
2    /* State */
3    PARTITION partition;
4    SEQNR next;
5    WINDOWS<REQUESTSOURCEID, REQUESTNR, REQUEST> requests;
6    WINDOWOPINIONS<COMMITTERID, RECORD> records;
7    SET<COMMITTERID> recorded;
8    OBSERVER<GCSOURCEID> gc;
9    OBSERVER<VIEWSOURCEID> view;
10   MODE mode;                          /* NORMAL, VIEW_CHANGE, or IDLE */
11
12   /* Initialization */
13   PROPOSER(PROPOSERID i, PARTITION p) : PROCESSOR(i) {
14     partition := p;
15     next := 0;
16     requests := WINDOWS();
17     records := WINDOWOPINIONS(0, WINDOW_SIZE);
18     recorded := SET();
19     gc := OBSERVER();
20     view := OBSERVER();
21     mode := ite(id == PROPOSER_OF_VIEW(view.current), NORMAL, IDLE);
22   }
23
24   /* Request handling */
25   VOID process(REQUESTTUPLE rt) {
26     If (requests[rt.creator].appendable(rt.request.rnr) == false) {
27       If (mode == NORMAL) return;
28       /* Make room for new request on idle proposers */
29       requests[r.source].move(r.rnr);
30     }
31     requests[rt.creator].append(rt.request);
32     If (mode == NORMAL) propose();
33   }
34
35   VOID propose() {
36     While (next < (gc.current + WINDOW_SIZE)) {
37       /* Determine request */
38       REQUEST r := nil;
39       For each REQUESTSOURCEID rqs in random order {
40         If (requests[rqs].min == requests[rqs].pos) continue;
41         r := requests[rqs][requests[rqs].min];
42         break;
43       }
44       If (r == nil) break;
45
46       /* Publish proposal */
47       CONSENSUSTUPLE proposal :=
48            CONSENSUSTUPLE(id, partition, next, view.current, r);
48       emit(proposal, COMMITTERS(partition));
49
50       /* Update state */
51       next := next + 1;
52       requests[r.source].move(r.rnr + 1);
53     }
54   }
55
56   /* Garbage collection */
57   VOID process(GCTUPLE gt) {
58     /* Update garbage-collection information */
59     If (gc.update(gt.creator, gt.snr()) == false) return;
60
61     /* Collect garbage */
62     records.move(gc.current);
63     If (mode == NORMAL) propose();
64   }
65
66   /* View change */
67   VOID process(VIEWTUPLE vt) {
68     /* Update view information */
69     If (view.update(vt.creator, vt.view) == false) return;
70
71     /* Trigger view change */
72     next := gc.current;
73     records.clear(next);
74     recorded := SET();
75     mode := ite(id == PROPOSER_OF_VIEW(view.current), VIEW_CHANGE,
                   IDLE);
76   }
77
78   VOID process(RECORDTUPLE rt) {
79     /* Check records */
80     If (mode != VIEW_CHANGE) return;
81     If (rt.view != view.current) return;
82     If (rt.creator ∈ recorded) return;
83
84     /* Store records */
85     next := max(next, rt.records.min);
86     records.move(rt.records.min);
87     For each SEQNR s in [rt.records.min, rt.records.pos]:
88          records[s] := rt.records[s];
```

```
88       recorded.add(rt.creator);
89       gc.current := max(gc.current, rt.records.min);
90
91       /* Return if the record information is not yet complete */
92       If (|recorded| < F+1) return;
93
94       /* Perform view change */
95       For each SEQNR s in [gc.current, gc.current + WINDOW_SIZE] {
96         /* Determine proposal */
97         RECORD r := RECORD(−1, nil);
98         For each COMMITTERID cmr {
99           If (s <= records[cmr].pos) continue;
100          If (records[cmr][s].view > r.view) r := records[cmr][s];
101        }
102        next := s;
103        If (r.view == −1) break;
104        requests[r.request.source].move(r.request.rnr + 1);
105
106        /* Publish proposal */
107        CONSENSUSTUPLE proposal := CONSENSUSTUPLE(id, partition,
                next, view.current, r.request);
108        emit(proposal, COMMITTERS(partition));
109      }
110
111      /* Switch mode */
112      mode := NORMAL;
113      propose();
114    }
115  }
```

## Committer

```
1  class COMMITTER extends PROCESSOR {
2    /* State */
3    PARTITION partition;
4    WINDOW<REQUEST> commits;
5    WINDOW<RECORD> records;
6    OBSERVER<GCSOURCEID> gc;
7    OBSERVER<VIEWSOURCEID> view;
8
9    /* Initialization */
10   COMMITTER(COMMITTERID i, PARTITION p) : PROCESSOR(i) {
11     partition := p;
12     commits := WINDOW(0, WINDOW_SIZE);
13     records := WINDOW(0, WINDOW_SIZE);
14     gc := OBSERVER();
15     view := OBSERVER();
16   }
17
18   /* Request handling */
19   VOID process(CONSENSUSTUPLE ct) {
20     /* Check proposal */
21     If (ct.view != view.current) return;
22     If (commits.appendable(ct.snr) == false) return;
23
24     /* Update state */
25     commits.append(ct.request);
26
27     /* Publish commit */
28     CONSENSUSTUPLE commit :=
            CONSENSUSTUPLE(id, partition, ct.snr, ct.view, ct.request);
29     emit(commit, EXECUTORS());
30   }
31
32   /* Garbage collection */
33   VOID process(GCTUPLE gt) {
34     /* Process tuple */
35     If (gc.update(gt.creator, gt.snr()) == false) return;
36
37     /* Collect garbage */
38     commits.move(gc.current);
39     records.move(gc.current);
40   }
41
42   /* View change */
43   VOID process(VIEWTUPLE vt) {
44     /* Update view information */
45     NUMBER old := view.current;
46     If (view.update(vt.creator, vt.view) == false) return;
47
48     /* Perform view change by updating records */
49     For each SEQNR s in [commits.min, commits.pos] {
50       records[s] := RECORD(old, commits[s]);
51     }
52     commits.reset();
53
54     /* Publish records */
55     RECORDTUPLE record :=
            RECORDTUPLE(id, partition, view.current, records);
56     emit(record, RECORD_SINKS(partition));
57   }
58  }
```

## Executor

```
1  class EXECUTOR extends PROCESSOR {
2    /* State */
3    APPLICATION application;
4    EXECUTIONNR next;
5    EXECUTORPARTITION[] partitions;
6    MAP<CLIENTID, COMMANDNR> filter;
7    MAP<CLIENTID, RESULT> results;
8    CHECKPOINTNR checkpoint;
9    OBSERVER<GCSOURCEID> gce;
10
11   /* Initialization */
12   EXECUTOR (EXECUTORID i) : PROCESSOR(i) {
13     application := APPLICATION();
14     next := 0;
15     partitions := EXECUTORPARTITION[#PARTITIONS];
16     filter := MAP();
17     results := MAP();
18     checkpoint := 0;
19     gce := OBSERVER();
20   }
21
22   /* Request handling */
23   VOID process (CONSENSUSTUPLE ct) {
24     /* Process tuple */
25     If (partitions[ct.partition].command(ct) == false) return;
26
27     /* Process commands */
28     While (true) {
29       /* Get command */
30       EXECUTORPARTITION p := partitions[next mod #PARTITIONS];
31       COMMAND x := p.commands[next / #PARTITIONS];
32       If (x == nil) break;
33
34       /* Execute command if necessary */
35       If (filter[x.xid] <= x.xnr) {
36         ANY a := application.execute(x.payload);
37         RESULT r := RESULT(x.xid, x.xnr, a);
38         results[x.xid] := r;
39         filter[x.xid] := x.xnr + 1;
40       }
41       next := next + 1;
42
43       /* Publish result if available */
44       If (results[x.xid].xnr == x.xnr) {
45         RESULTTUPLE result := RESULTTUPLE(id, results[x.xid]);
46         emit(result, RESULT_SINKS(partition));
47       }
48
49       /* Create snapshot if necessary */
50       CHECKPOINTNR c := next / CP_INTERVAL;
51       If (checkpoint < c) {
52         /* Create and store snapshot */
53         SNAPSHOT snapshot := SNAPSHOT(application.state, filter, results);
54         Store snapshot at SNAPSHOT_LOCATION(id, c);
55         checkpoint := c;
56
57         /* Publish new checkpoint */
58         GCTUPLE checkpoint := GCTUPLE(id, c, nil);
59         emit(checkpoint, GC_SINKS());
60       }
61     }
62   }
63
64   /* Garbage collection */
65   VOID process (GCTUPLE gt) {
66     /* Process tuple */
67     If (gce.update(gt.creator, gt.cnr * CP_INTERVAL) == false) return;
68
69     /* Load snapshot if necessary */
70     CHECKPOINTNR cnr := gce.current / CP_INTERVAL;
71     If (next < gce.current) {
72       /* Get snapshot */
73       SNAPSHOT snapshot := nil;
74       For each EXECUTORID e in gt.executors {
75         snapshot := Try to load snapshot from SNAPSHOT_LOCATION(e, cnr);
76         If (snapshot != nil) break;
77       }
78       If (snapshot == nil) return;
79
80       /* Update state */
81       application.state := snapshot.state;
82       next := cnr * CP_INTERVAL;
83       For each PARTITION p: partitions[p].gc(next / #PARTITIONS);
84       filter := snapshot.filter;
85       results := snapshot.results;
86       checkpoint := cnr;
87       Store snapshot at SNAPSHOT_LOCATION(id, cnr);
88     } else {
89       /* Move windows */
90       For each PARTITION p: partitions[p].gc(gce.current / #PARTITIONS);
91     }
92
93     /* Discard old snapshots */
94     Discard all snapshots with CHECKPOINTNR c < cnr;
95   }
96
97   /* View change */
98   VOID process (VIEWTUPLE vt) {
99     partitions[vt.partition].view(vt);
100  }
101
102  Periodically {
103    For each PARTITION p {
104      PROGRESSTUPLE pt := PROGRESSTUPLE(id, partitions[p].progress);
105      emit(pt, CONTROLLERS(p));
106    }
107  }
108 }
```

## Executor Partition (Auxiliary Class)

```
1  class EXECUTORPARTITION {
2    /* State */
3    WINDOWOPINIONS<COMMITTERID, REQUEST> commits;
4    WINDOW<COMMAND> commands;
5    PROGRESS progress;
6    OBSERVER<VIEWSOURCEID> view;
7
8    /* Initialization */
9    EXECUTORPARTITION () {
10     commits := WINDOWOPINIONS(0, WINDOW_SIZE);
11     commands := WINDOW(0, WINDOW_SIZE);
12     progress := PROGRESS();
13     view := OBSERVER();
14   }
15
16   /* Request handling */
17   BOOLEAN command (CONSENSUSTUPLE ct) {
18     /* Check tuple */
19     If (ct.view != view.current) return false;
20     If (commits[ct.creator].appendable(ct.snr) == false) return false;
21
22     /* Update state */
23     commits[ct.creator].append(ct.request);
24     If (commits.available(ct.snr) < F+1) return false;
25
26     /* Consensus process is complete */
27     commits.fill(ct.snr + 1);
28     commands[ct.snr] := ct.request.command;
29     progress[ct.request.source] := ct.request.rnr + 1;
30     return true;
31   }
32
33   /* Garbage collection */
34   VOID gc (SEQNR s) {
35     commits.move(s);
36     commands.move(s);
37   }
38
39   /* View change */
40   VOID view (VIEWTUPLE vt) {
41     /* Update view state */
42     If (view.update(vt.creator, vt.view) == false) return;
43
44     /* Handle view change */
45     commits.sync(commands.pos);
46   }
47 }
```

```
1  class CONTROLLER extends PROCESSOR {
2      /* State */
3      PARTITION partition;
4      MAP<REQUESTSOURCEID, TIMESTAMP> reports;
5      PROGRESSOPINIONS<EXECUTORID, PROGRESS> ordered;
6      PROGRESS target;
7      PROGRESS actual;
8      TIMEOUT timeout;
9      TIMESTAMP deadline;
10     OBSERVER<VIEWSOURCEID> view;
11     MODE mode;                                        /* NORMAL or IDLE */
12
13     /* Initialization */
14     CONTROLLER(CONTROLLERID i, PARTITION p) : PROCESSOR(i) {
15         partition := p;
16         reports := MAP();
17         For each REQUESTSOURCEID rqs in
               REQUEST_SOURCES(partition): reports[rqs] := now;
18         ordered := PROGRESSOPINIONS();
19         target := PROGRESS();
20         actual := PROGRESS();
21         timeout := CONTROLLER_TIMEOUT;
22         deadline := ∞;
23         view := OBSERVER();
24         mode := NORMAL;
25     }
26
27     /* Progress monitoring */
28     VOID process(PROGRESSTUPLE pt) {
29         /* Check mode */
30         If (mode == IDLE) return;
31
32         /* Update progress information */
33         If (pt.creator is a REQUESTSOURCEID) {
34             If (pt.progress[pt.creator] <= target[pt.creator]) return;
35             target[pt.creator] := pt.progress[pt.creator];
36             reports[pt.creator] := now;
37         } else if (pt.creator is a EXECUTORID) {
38             If (pt.progress ⪯ ordered[pt.creator]) return;
39             ordered[pt.creator] := pt.progress;
40             PROGRESS p := ordered.highest(F+1);
41             If (actual ≺ p) timeout := CONTROLLER_TIMEOUT;
42             actual := p;
43         }
44
45         /* Update deadline */
46         deadline();
47     }
48
49     VOID deadline() {
50         TIMESTAMP time := ∞;
51         For each REQUESTSOURCEID rqs in REQUEST_SOURCES(partition) {
52             If (actual[rqs] < target[rqs]) time := min(time, reports[rqs] +
                   timeout);
53         }
54         deadline := time;
55     }
56
57     /* View change */
58     VOID process(VIEWTUPLE vt) {
59         /* Update view information */
60         If (view.update(vt.creator, vt.view) == false) return;
61
62         /* Enable monitoring */
63         target := PROGRESS();
64         mode := NORMAL;
65
66         /* Update deadline */
67         deadline();
68     }
69
70     Periodically {
71         /* Check deadline */
72         If (mode == IDLE) return;
73         If (now < deadline) return;
74
75         /* Disable monitoring */
76         timeout := timeout * 2;
77         mode := IDLE;
78
79         /* Publish new view */
80         VIEWTUPLE view := VIEWTUPLE(id, partition, view.current + 1);
81         emit(view, VIEW_SINKS(partition));
82     }
83 }
```



\end{document}